\documentclass{article}

\usepackage{microtype}
\usepackage{graphicx}
\usepackage{subfigure}
\usepackage{booktabs} 
\usepackage{soul}

\usepackage{hyperref}

\usepackage[accepted]{icml2024}

\usepackage{amsmath}
\usepackage{amssymb}
\usepackage{mathtools}
\usepackage{amsthm}

\usepackage[capitalize,noabbrev]{cleveref}

\theoremstyle{plain}

\theoremstyle{definition}

\theoremstyle{remark}

\usepackage[textsize=tiny]{todonotes}

\icmltitlerunning{The Scaling Law in Stellar Light Curves}

\begin{document}

\twocolumn[
\icmltitle{The Scaling Law in Stellar Light Curves}



\icmlsetsymbol{equal}{*}

\begin{icmlauthorlist}
\icmlauthor{Jia-Shu Pan}{equal,ucas,rsaa,naoc}
\icmlauthor{Yuan-Sen Ting}{equal,rsaa,soc,osu}
\icmlauthor{Yang Huang}{ucas,naoc}
\icmlauthor{Jie Yu}{rsaa,mpi,hits}
\icmlauthor{Ji-Feng Liu}{naoc,ucas}

\end{icmlauthorlist}

\icmlaffiliation{rsaa}{Research School of Astronomy \& Astrophysics, Australian National University, Cotter Rd., Weston, ACT 2611, Australia}
\icmlaffiliation{ucas}{School of Astronomy and Space Science, University of Chinese Academy of Sciences, Beijing 100049, China}
\icmlaffiliation{naoc}{CAS Key Lab of Optical Astronomy, National Astronomical Observatories, Chinese Academy of Sciences, Beijing 100012, China}
\icmlaffiliation{soc}{School of Computing, Australian National University, Acton, ACT 2601, Australia}
\icmlaffiliation{osu}{Department of Astronomy, The Ohio State University, 140 West 18th Avenue, Columbus, OH 43210, USA}
\icmlaffiliation{mpi}{Max Planck Institute for Solar System Research, Justus-von-Liebig-Weg 3, 37077 Göttingen, Germany}
\icmlaffiliation{hits}{Heidelberg Institute for Theoretical Studies (HITS) gGmbH, Schloss-Wolfsbrunnenweg 35, D-69118 Heidelberg, Germany}

\icmlcorrespondingauthor{Jia-Shu Pan}{jspan@smail.nju.edu.cn}

\icmlkeywords{Machine Learning, ICML}

\vskip 0.3in
]



\printAffiliationsAndNotice{\icmlEqualContribution} 

\begin{abstract}
Analyzing time series of fluxes from stars, known as stellar light curves, can reveal valuable information about stellar properties. However, most current methods rely on extracting summary statistics, and studies using deep learning have been limited to supervised approaches. In this research, we investigate the scaling law properties that emerge when learning from astronomical time series data using self-supervised techniques. By employing the GPT-2 architecture, we show the learned representation improves as the number of parameters increases from $10^4$ to $10^9$, with no signs of performance plateauing. We demonstrate that a self-supervised Transformer model achieves 3-10 times the sample efficiency compared to the state-of-the-art supervised learning model when inferring the surface gravity of stars as a downstream task. Our research lays the groundwork for analyzing stellar light curves by examining them through large-scale auto-regressive generative models.
\end{abstract}

\section{Introduction}
\label{sec:intro}

Light curves, temporal brightness variations of celestial objects, can reveal much of the astrophysics of these systems and aid in the discovery of fleeting transient phenomena \cite{astero}. This realization has prompted a rapidly growing field in the last decade, with a myriad of missions such as \textit{Kepler} \cite{kepler}, TESS \cite{tess}, and ZTF \cite{ztf} drastically transforming the landscape of modern-day astronomy. This trend is further accelerating, with upcoming surveys such as the Rubin Observatory \cite{lsst} and SiTian \cite{sitian} aiming to collect light curves from up to 10 billion objects from the sky within this decade.

Despite the flood of time series data in astronomy, characterizing these light curves to perform downstream label inferences or identify out-of-distribution objects remains an unresolved problem. Traditional methods often rely on crude summary statistics \cite{flicker}. For example, in the field of asteroseismology, which involves characterizing light curves from stars, the established method calculates the power spectrum of the light curves and extracts the frequency of the maximum oscillation power, $\nu_{\rm max}$ , which highly correlates with stellar masses and ages \cite{astero}. Therefore, asteroseismology has been the primary way to achieve golden measurements of how old stars are. However, such summary statistics do not fully harness all the information contained in the light curves and are often difficult to determine due to heterogeneous noise levels \cite{tess_astero} and irregular cadences \cite{lsst_astero}.

To overcome this limitation, a growing body of literature has developed machine learning methods for time-domain astronomy \cite{cnn_lc, astroconformer}. In particular, for stellar light curves, Transformer models \cite{transformer} is receiving an increasing interest \cite{astromer}. For example, \citet{astroconformer} customized Transformer models for light curves and demonstrated that, in the case of extracting the surface gravity (or $\log g$) of stars, the ability of Transformer models to extract long-range information can yield state-of-the-art performance in a supervised learning setting.

While deep learning has shown promise in characterizing stellar light curves, most studies have been limited to discriminative supervised learning \cite{cnn_ps, swan}. Such methods often fall short in applications due to two main reasons. First, there is a lack of existing ground truth labels for the data. For example, while the Rubin Observatory will collect tens of billions of time series data, the number of existing labels (obtained from other more observationally expensive means -- such as monitoring eclipsing binaries) will remain on the order of thousands \cite{binary_kepler}. Second, there is a need for a generative model that can extract robust, generalizable representations suitable for other downstream tasks and out-of-distribution search. And solving these bottlenecks is critical to limit redundant resource dedications for different pipelines \cite{cnn_ps, swan, astroconformer} and downstream tasks.

This limitation has called for a different approach that can utilize unlabeled data and can generalise for various downstream tasks. A prominent candidate would be training foundational models through self-supervised learning as such models have shown promises in different domains such as text data and images \cite{gpt-3, foundation_image}. These models, especially Large Language Models (LLMs), exhibit performance unmatched by their smaller variants and emergent abilities such as in-context learning \cite{gpt-3}. Furthermore, in recent years, foundational models in other domains has observed the scaling law, which is the empirical observation that the next-token prediction loss of Transformer-based autoregressive generative models scales in a power law with increasing compute budgets, model size and the number of training tokens \cite{scaling_lm, scaling_ar, scaling_chinchilla}. Applying the same idea to astronomical time series data invites two questions: (1) Would the same self-supervised generative models based on Transformer exhibit the scaling law on astronomical time series data? (2)  Would a better next-token prediction loss obtained by scaling up translate to better performance in downstream tasks of interest?

\section{\textit{Kepler} Mission's Stellar Light Curves}
\label{sec:data}

In this study, we focus on the stellar light curves from the \textit{Kepler} mission. The \textit{Kepler} mission was a flagship NASA mission that operated from 2009 to 2013. It yielded 4 years of light curves for approximately 200,000 stars at an equal cadence of about 29.4 minutes, containing $\sim$70,000 timestamps and corresponding observed fluxes. In order to compare with the supervised deep learning Transformer-based models \cite{astroconformer}, we follow the same \textit{Kepler} data selection, selecting 17,201 high-quality light curves as our dataset. Some light curve examples are shown in Fig 1. The different panels show the cases with four representative stars with different surface gravity values, where a smaller $\log g$ indicates a larger stellar radius and, consequently, less oscillation in the stellar fluxes.

The light curves are selected based on the availability of the asteroseismic $\log g$ values for these stars, as provided in previous studies \cite{mathur, yu18, yu20}. These studies focus on extracting the properties of stars with the maximum oscillation frequency $\nu_{\rm max}$, and the prominence of $\nu_{\rm max}$ makes it an effective way to select light curves that have high signal-to-noise ratios. Furthermore, the radii of the stars are also provided in \citet{berger}, which allow for the application of a radius-dependent high-pass filter to remove instrumental effects \cite{prep_kepler}.

After filtering and outlier removal, we segment the light curves into non-overlapping adjacent observations in patches of length 80, with each patch representing a single sample in our training batch. We use every single timestamp as an input token (see Section~\ref{sec:method}), amounting to 0.7B tokens. We chose a relatively short context window (containing only 80 tokens) because \textit{Kepler} observations are not contiguous and have observational time gaps between the data points. For simplicity, we opted for these short context windows to ensure that all light curves are contiguous without the distraction of time gaps. It is possible that the autoregressive generative models studied here would remain robust even with the presence of time gaps, but we chose to leave this investigation for future study, as the primary goal of this work is to demonstrate the scaling law of the autoregressive generative models. Finally, our dataset is split into a training set and a test set with a ratio of 25:1.

\section{Learning Stellar Light Curves with GPT-2}
\label{sec:method}

In this study, we adopt the GPT-2 architecture \cite{gpt-2} to investigate the scaling law for astronomical stellar light curves. We chose the GPT-2 because, unlike some of the later variants like Llama \cite{llama} and Qwen \cite{qwen}, GPT-2 provides a more principled and simple architecture that allows us to easily scale the models with different numbers of parameters and with minimal ambiguity. The model has demonstrated its effectiveness and transferability in text \cite{gpt-2}, image \cite{iGPT}  and beyond in galaxy images\footnote{Previously, to the best of our knowledge, scaling laws have only been established in supervised learning in astronomy \cite{scaling_sup}. Complementary to this paper, as our work was being developed, we recognize that a similar study was independently conducted for galaxy images and released around the same time as ours \cite{astropt}.}. 


GPT-2 is a decoder-only Transformer with learnable and straightforward positional embeddings and Pre-layer Normalization, in contrast to more recent LLM architecture such as Llama \cite{llama} and Qwen \cite{qwen}, which instead use Rotary Positional Embedding \cite{rope} and RMSNorm \cite{rmsnorm}. We train all our models following the training scheme in \citet{gpt-2}. We adopt the AdamW optimizer \cite{adamw}. The learning rate is increased linearly to the peak value in 2,000 iterations and then annealed to $2\times10^{-4}$ with a cosine schedule while ensuring the learning rate remains higher than $2\times10^{-5}$. We also adopt the default batch size of 12 and accumulate gradients for 40 batches, leading to an effective batch size of 480 light curve patches or 38,400 tokens. All models are trained for 200,000 iterations, summing up to 7.7B processed tokens.

For the GPT-2 model to work with time series data, some minor modifications are required. In particular, unlike discrete variable sequences such as texts, where the modeling of conditional distribution can be handled by minimizing the KL divergence between the text distribution and the output categorical distribution, continuous distribution is less straightforward to model. Instead, we carry out next-token prediction by regressing the next token with Huber loss \cite{huber}, a rectified version of mean square error (MSE) that mitigates the influence of outlier values. Note that this is equivalent to modeling the expectation of the conditional distribution, or performing maximum likelihood estimation with the assumption that the conditional distribution is a constant-variance normal distribution. 

Furthermore, the input of GPT-2 models is in the form of tokens from text data. To adapt it to light curves, we replace the token embedding layer and language modeling layer in the model with two learnable Multilayer Perceptrons (MLPs) whose dimension is listed in Table~\ref{tab:model}. 

\begin{table}[t]
\caption{The different autoregressive generative models investigated in this study. We study the four GPT-2 models (default, medium, large, and XL) as proposed in \cite{gpt-2}. We also vary the default GPT-2 model by changing the the depth (number of layers), head (number of self-attention heads), as well as width (hidden dimension size) of the Transformer architecture, to allow us to test a wider dynamic range in terms of model complexity.}
\label{tab:model}
\vskip 0.15in
\begin{center}
\begin{tabular}{lcccr}
\toprule
Parameter (Million) & Depth & Head & Width \\
\midrule
0.04  & 3& 4 & 32 \\
0.15  & 3& 4 & 64 \\
0.59  & 3& 4 & 128 \\
2.36 & 3& 4 & 256 \\
9.44  & 3& 8 & 512  \\
37.8      & 12& 8& 512    \\
85 (GPT-2, default)     & 12& 12& 768\\
308 (medium)   & 24& 16&  1024   \\
708 (large)   & 36& 20&   1280     \\
1477 (XL)  & 48& 25& 1600 \\
\bottomrule
\end{tabular}
\end{center}
\vskip -0.1in
\end{table}

In addition to the original GPT-2, GPT-2 medium, GPT-2 large, and GPT-2 XL models proposed in \citet{gpt-2}, we also train models of different sizes by adjusting the number of layers (depth), number of self-attention heads (Head), and hidden dimension size (width), as listed in Table~\ref{tab:model}. This is to ensure that we can evaluate the performance of the models with a larger dynamic range in terms of model size, ranging from $10^4$ to $10^9$ parameters. We note that the largest model that we train, with $\sim 1.5$B parameters, starts to rival some of the ``lightweight" modern-day LLMs \cite{qwen}.

\begin{figure}[ht]
\vskip 0.2in
\begin{center}
\centerline{\includegraphics[width=\columnwidth]{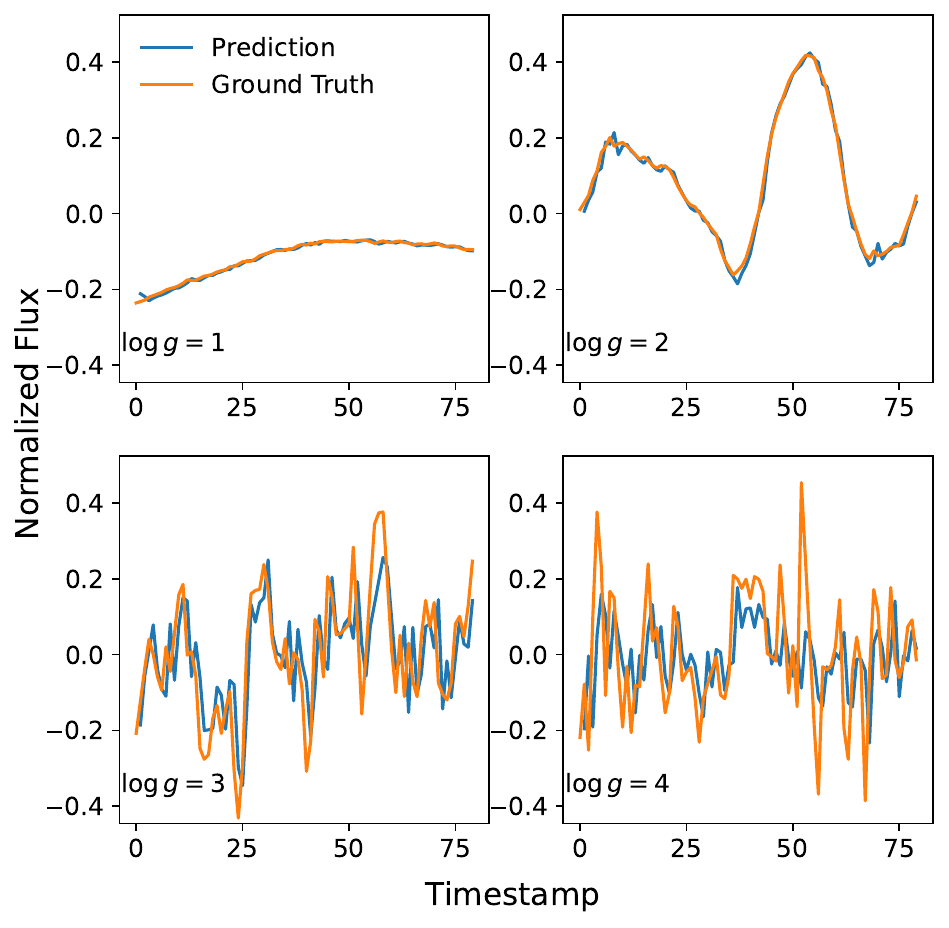}}
\caption{Autoregressive one-step predictions from our GPT-2 XL model with 1.5B parameters. The different panels show four representative light curves with varying surface gravity ($\log g$) values of stars. We perform the next-step prediction conditioning on all the previous $N$ steps. The generative model demonstrates the ability to capture the general trend of the light curves, leading to a robust representation of the light curves.
}
\label{fig:prediction_logg}
\end{center}
\vskip -0.2in
\end{figure}

\section{Results}
\label{sec:results}

The study of \textit{Kepler}'s light curves is a cornerstone in modern-day stellar astrophysics. The temporal variations of the fluxes of stars are the manifestation of physical processes where light bounces within the stellar interior. As such, the light curves are tell-tale signatures of the stellar interior physics that are otherwise inaccessible to us. In this study, we train GPT-2-like models as autoregressive generative models to capture this inherent interior physics. Note that, the models are all initialized completely randomly from $\mathcal N(0, 0.02)$ to study the emergence of the scaling law in the generative models trained from scratch.

We will start by showcasing the qualitative results from the GPT-2 XL model with $1.5$B parameters. In Figure~\ref{fig:prediction_logg}, we show the one-step predictions of all fluxes, where at each step, we perform the prediction conditioning on all the previous (true) observed fluxes. We note that, given the predominantly stochastic nature of stellar flux variations, i.e., given all $N$ previous steps, the next ``token" prediction complies to a distribution, just like in natural languages. As such, we do not expect the predictions to be exact. Specifically, since our prediction is based upon maximal likelihood estimation from a multi-mode distribution, the multi-step prediction would deviate from the truly observed fluxes although the predicted light curve could be statistically the same as the original light curves. Due to this consideration, we only examine the trend with one-step predictions.

As shown in Figure~\ref{fig:prediction_logg}, like in natural languages, the autoregressive generative model does learn the ``stylistic patterns" of the stellar light curves, where it shows a good understanding of the long timescale oscillations from stars with small surface gravity as opposed to short oscillations from stars with large surface gravity. More importantly, the ability to generally grasp the patterns of the light curves suggests that we could further improve the predictions if a scaling law applies to the model, and the learned latent representations might be robust for further downstream tasks such as predicting the surface gravity (and hence ages) of the stars.

\begin{figure*}[ht]
\vskip 0.2in
\begin{center}
\centerline{\includegraphics[scale=0.8]{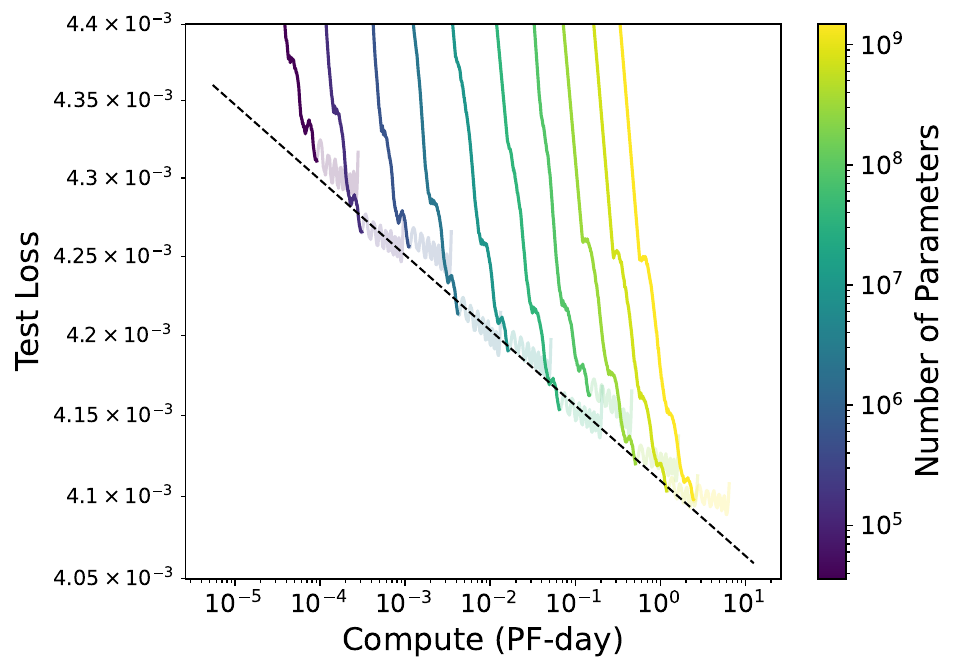}}
\caption{The emergence of scaling law for stellar light curves. We train autoregressive generative models on stellar light curves with different model sizes, ranging from $10^4$ to $10^9$ parameters. The different lines show the pretraining loss (Huber loss of next-token prediction) as a function of the compute budget for different model sizes. We truncate the training of the models when the slope of the tangent line connecting the loss from the last three epochs becomes shallower than the overall scaling law. The prediction loss plateaus at lower values for larger models, demonstrating that the scaling law also applies to Transformer-based autoregressive generative models when applied to stellar light curves.}
\label{fig:loss_flops}
\end{center}
\vskip -0.2in
\end{figure*}

\subsection{The Emergence of Scaling Law for Light Curves}
\label{sec:scaling_law}

While the autoregressive model shows some promise, a key to understanding if astronomical time series data can benefit from a foundational model would be determining if the same scaling law that has been observed in real-life applications also translates to stellar light curves. Such a study has not been established before, and it is also not immediately clear if that would necessarily be the case. 

Unlike text data, the sample size of astronomical time series adopted in this study is much smaller. With the current data described in Section~\ref{sec:data}, the high-quality \textit{Kepler} light curves in this study only have 0.7B tokens. Even if we scale it up to the entire \textit{Kepler} catalog, it would only lead to an order of magnitude larger in terms of training data. This is minuscule compared to modern-day LLMs, which are typically trained on trillions of tokens. But this limitation in data might be compensated by the fact that, the underlying structure of stellar light curves is also much simpler than natural languages, with only finite dynamos driving the stochastic variations of fluxes.

We investigate the scaling law between pretraining loss and compute budget with model sizes spanning $10^4-10^{9}$ parameters. Figure~\ref{fig:loss_flops} shows the pretraining loss of the test set as a function of the compute budget in PetaFlop(PF)-days\footnote{1 PF-day = $10^{15}\times24\times3600=8.64\times10^{19}$ floating point operations $\approx 3.2$ A100 days.} for different GPT-2-like models, with individual lines of different colors representing the test loss curve for various compute budgets. We define the critical point as 3.4 epoch, where the local slope of the loss curve becomes shallower than the overall scaling law defined below. The models are trained for 11.4 epochs. The loss curves before the critical point are shown in solid colors, while the curves after the critical points are depicted with more translucent colors.

Following \citet{scaling_ar}, we fit a power law to the loss at the critical points as a function\footnote{The original expression in \citet{scaling_ar} is $L(C) = (\frac{C_0}{C})^{\alpha_C}$, we apply a  base-10 logarithm to the expression for better readability. Throughout this study, we consistently utilize the modified equivalent expression.} of the compute budget $C$,  $\log_{10} L(C) =\alpha_C \log_{10} C + \beta_C$  with $\alpha_C=-0.00448, \beta_C=-2.386$, demonstrating the emergence of a scaling law in stellar light curve data. The scaling suggests that, for a 10 fold larger compute budget, the next-token prediction Huber loss (again, a rectified version of MSE) would improve by 1\%. Already, we can predict that next-token prediction Huber loss will improve 3.8\% compared to our most capable 1.5B GPT-2 XL model at a compute budget of $10^4$ PF-days.

We suspect the somewhat shallow slope, i.e., smaller gain as model size increases, is due to the limited training data size, as scaling laws observed in \cite{scaling_lm} are also functions of the dataset size. We aim to explore the dependency on the training set size in our future work.

\subsection{Scaling Law in Inferring Stellar Surface Gravity}
\label{sec:downstream}

While we have demonstrated the scaling law in the pretraining loss, in our case, the Huber loss of the next-token prediction, perhaps a more important question pertaining to most astronomical tasks is whether the latent representation acquired from such self-supervised learning also sufficiently exhibits the scaling law. The transferability and scalability to downstream tasks will ultimately determine if a foundational model for astronomical time series data is attainable.
\begin{figure}[ht]
\centerline{\includegraphics[width=0.85\columnwidth]{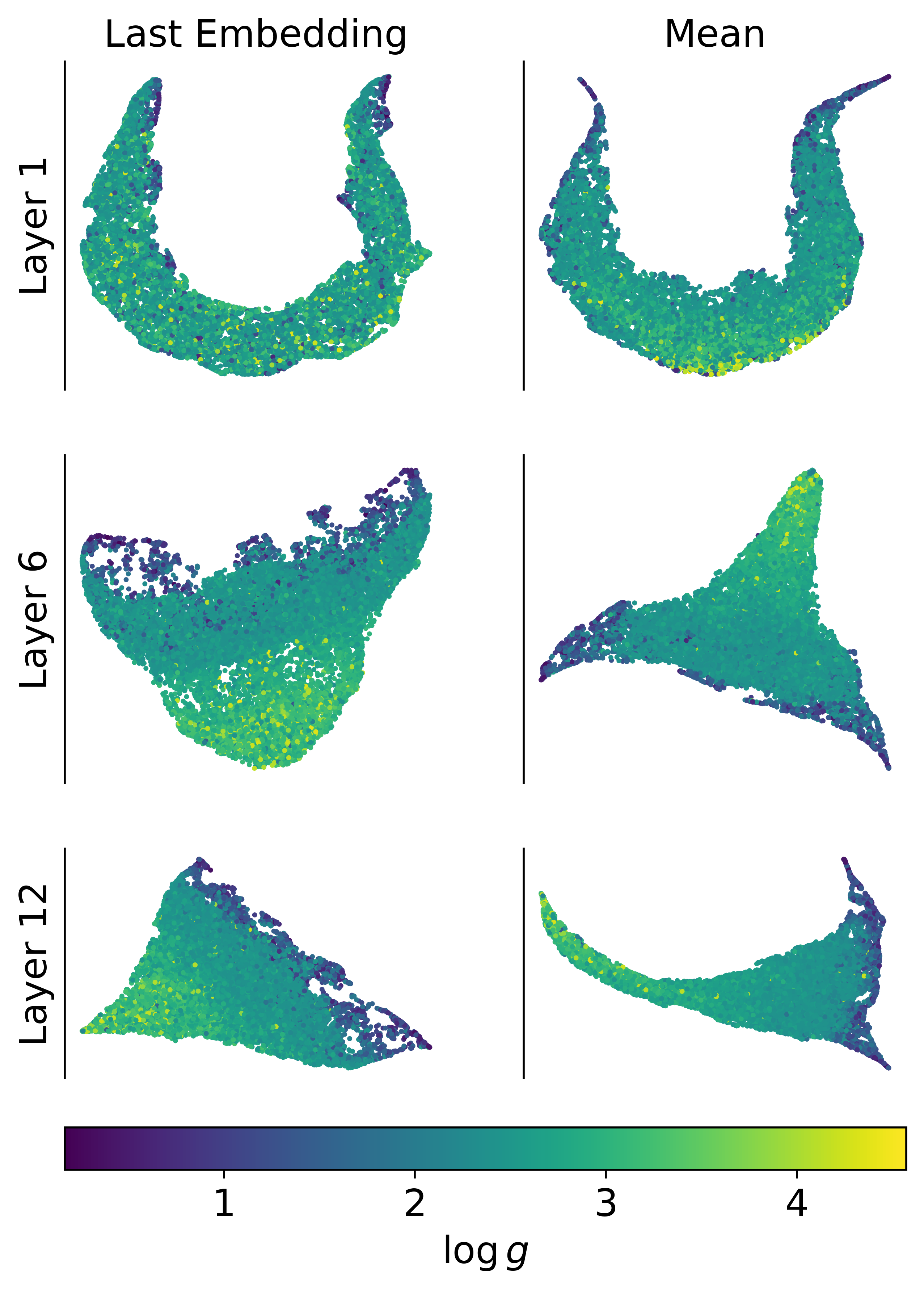}}
\caption{The latent representations extracted from the GPT-2 model. The native representation is a vector with 768 dimensions, which we subsequently visualize in 2D with the UMAP projection. Different columns show the latent embedding representation extracted at different depths in the generative model. The left panels show the case where only the last token is extracted as the embedding, while the right panels show the average of all 80 tokens at any given layer. Embeddings in the deeper layer show a higher level of abstraction. The symbols are color-coded by the surface gravity values of the stars.}
\label{fig:representation_layer}
\end{figure}

We first investigate how representation quality varies with model depth in the GPT-2 model.  For each light curve patch, we extract the embedding of the last token across all layers. The embeddings are then visualized with a UMAP \cite{umap} projection color-coded by their $\log g$ in Figure~\ref{fig:representation_layer}. As demonstrated in the figure, the representation at deeper layers shows a higher level of abstraction, leading to more distinct and discernible patterns with the $\log g$ labels. Since the radius, and hence $\log g$, of the stars is the primary factor that determines stellar oscillation (and consequently the emergent variation in the fluxes of stars), this gradual distinct representation shows that the learned representation has a high level of generalizability. We also compare with using only the last token's embedding in specific layers, as shown on the left panels, versus the one where we adopt an average of all tokens at the layer, as shown on the right panels, and found the latter yields cleaner representations. Hence, in the following, we will only adopt the latter for the downstream task.

For the label inference head (mapping from the last layer representation to $\log g$), we adopt a 3-layer MLP with a hidden dimension of 64. As different models have different width, the dimensions for the representation and hence the input dimension of the MLP head also varies. The largest MLP head has 0.11M parameters. We freeze the main generative model and train using a dataset consisting of 10,000 patches for training and 20,000 patches for validation. As mentioned in Section~\ref{sec:data}, our pretraining dataset contains over 8 million patches from more than 16,000 stars. The 10,000 patches used for the downstream task are selected from this pretraining dataset, corresponding to about $10,000$ stars. This sample size is representative of the number of golden labels obtained through other means in a practical setting. To ensure no information leakage, we use a separate test set of 20,000 patches drawn from the test set, which was not used during the pretraining of the generative models.

All our tests below for the label inference are on this unseen test set. For each model, we train 10 MLP heads, and the one with the best MSE loss on the validation set mentioned above is adopted as the final model. For the optimization, we use Adam optimizer \cite{adam} with a constant learning rate of 0.001 and adopt the full batch of the training data, as we find that stochastic gradient descent with minibatches does not alter the results, and the former is computationally more efficient. We set a maximum number of epochs to be 1000 and assume an early stop if the validation loss does not improve for over the last 50\% of the iterations.

\begin{figure}[t]
\begin{center}
\centerline{\includegraphics[width=\columnwidth]{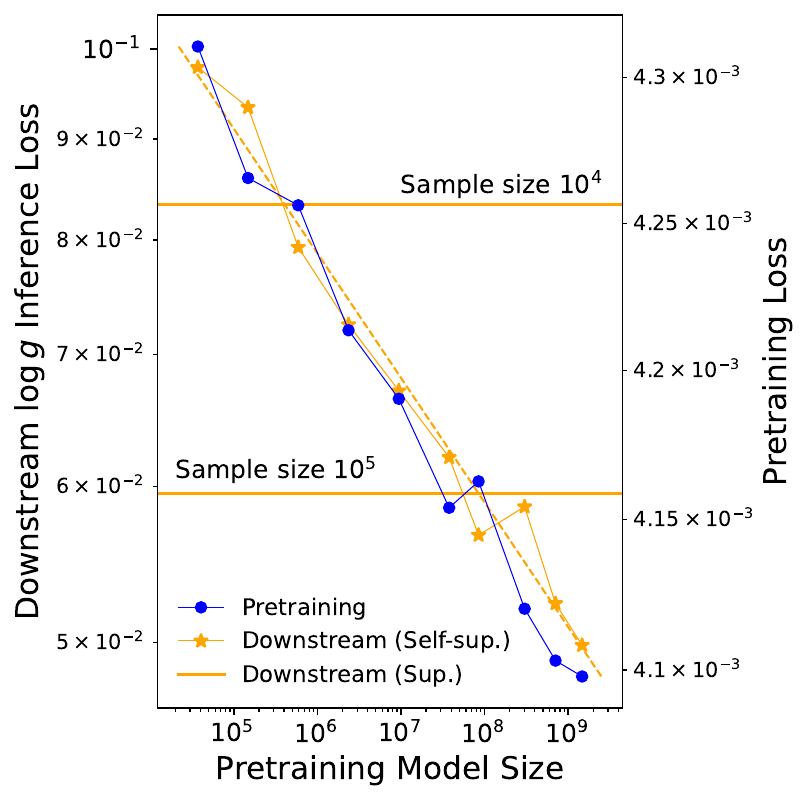}}
\caption{The downstream $\log g$ inference exhibits the scaling law. The solid orange line shows the MSE of the $\log g$ inference derived from the representation of the GPT-2 models through a final MLP head. The MSE is plotted as a function of the number of parameters. We also plot the Huber loss from the next-token prediction, as illustrated in Figure~\ref{fig:loss_flops}, as the blue solid line. The downstream $\log g$ prediction MSE closely follows the next-token prediction loss. We also compare the $\log g$ prediction with the Astroconformer models trained through supervised learning on 10,000 and 100,000 samples. We found that the generative approach in this study, at the $10^8$ parameter level, rivals the performance of the state-of-the-art supervised learning model but achieves this with 10 times fewer training data.}
\label{fig:logg_params}
\end{center}
\end{figure}

Figure~\ref{fig:logg_params} shows the precision of the $\log g$ inference as a function of the number of parameters, represented by the solid orange line. For reference, the next-token prediction loss, as previously discussed, is plotted in blue. The results demonstrate that the $\log g$ inference from the GPT-2-like model also exhibits a prominent scaling law with respect to the number of parameters in the model, and the improvement in this downstream label inference mirrors the next-token predictions. This suggests that the scaling law observed in the next-token prediction goes beyond simply memorizing the training set, and the model actually learns more robust abstractions from the light curves.

We compare our results from self-supervised learning with GPT-2 models to the current state-of-the-art (SOTA) supervised learning method, Astroconformer \cite{astroconformer}, represented by the solid orange horizontal lines. Astroconformer is a Transformer-based model with 0.86M parameters that learns from the direct relationship between the stellar light curves and $\log g$. They argue that introducing a strong inductive bias is necessary because the labeled training sample is limited. To address this, they combine convolution and self-attention mechanisms to simultaneously capture localized and global correlations, which they found to outperform other existing supervised learning methods primarily based on convolutional neural networks or K-nearest neighbor search \cite{swan}. To ensure a fair comparison, the Astroconformer results presented here are trained with the same labeled training samples used for evaluating self-supervised representations. Further, to explore the sample efficiency gain by self-supervised learning, we also evaluate Astroconformer's performance based on 10 times more training samples drawn from the pretraining dataset.

The figure shows that the scaling law enables sufficiently large self-supervised autoregressive models, even with the 0.7B pretraining tokens, to surpass the performance of the current supervised learning models. Astroconformer reaches a $\log g$ prediction MSE of 0.083 and 0.059 when trained on 10,000 samples and 100,000 samples, respectively. Even with the 85M parameters GPT-2 model, the learned latent representation is robust to surpass supervised learning trained on 10 times of data, which indicates they are $>10$ times more label-efficient. We fit the power law as before, and find the scaling law for the $\log g$ prediction MSE follows $\log_{10}L(N) = -0.063\log_{10} N - 0.725$. In other words, if we extrapolate that to the typical industrial scale of 10B or 100B, the $\log g$ precision MSE would be 11\% and 23\% lower than our most capable 1.5B GPT-2 XL model.

\begin{figure}
\vskip 0.2in
\begin{center}
\centerline{\includegraphics[width=\columnwidth]{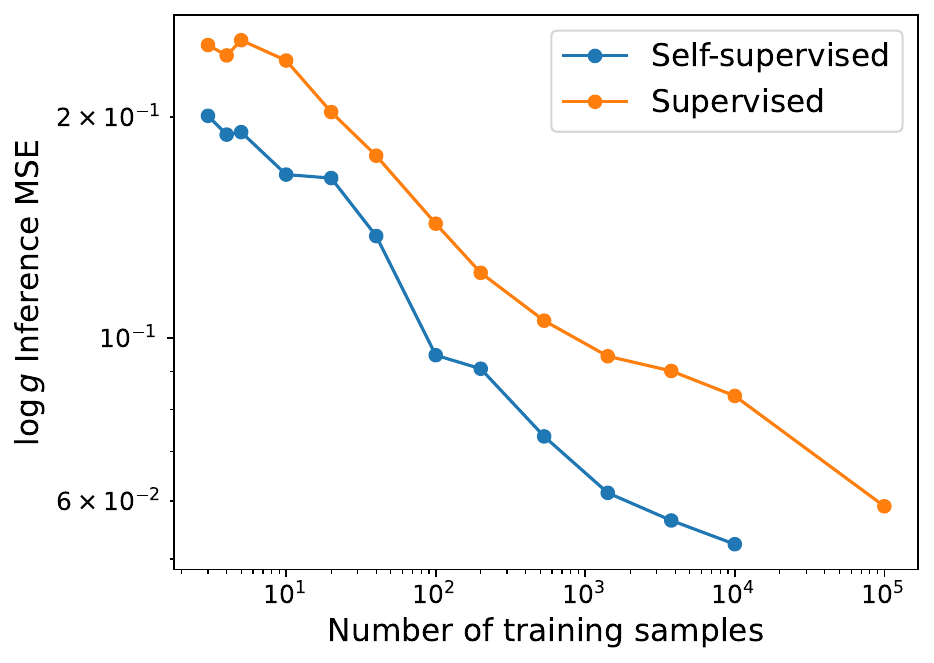}}
\caption{The impact of the labeled training dataset on the downstream $\log g$ inference task. We compare the performance of GPT-2 XL with that of Astroconformer. For any given fixed set of labeled data, the autoregressive generative approach adopted in this study leads to better precision than the supervised method. The self-supervised learning approach requires less than one third of training samples to achieve the same precision in $\log g$ inference.}
\label{fig:precision_label}
\end{center}
\vskip -0.2in
\end{figure}

In the comparison above, we assume a somewhat optimistic number of labeled samples (10,000) for downstream training. However, a key advantage of training a potentially large-scale foundation model is its ability to generalize with a small number of labeled samples. In Figure~\ref{fig:precision_label}, we explore how the downstream MSE loss varies as a function of the labeled dataset size using the GPT-2 XL model and contrast the results with Astroconformer's. As shown in Figure~\ref{fig:precision_label}, pretraining the autoregressive generative models typically leads to the same $\log g$ inference precision but with less than one third of the number of labeled samples. 


\section{Discussion and Conclusion}
\label{sec:discussion}

In this study, we explore training autoregressive generative models, based on the GPT-2 architecture, on stellar light curves. Stellar light curves have been an instrumental part of modern-day astrophysics to decipher our knowledge about the stellar interior and infer fundamental parameters of the stars. However, by nature, the number of stars with ground-truth labels obtained through other observational means is highly limited. As such, a scalable deep learning approach based on self-supervision, a method that has shown promise primarily in text data, but now also in other modalities, is critical to move the field of asteroseismology forward.

We demonstrate that with stellar light curves, there is also an emergence of a scaling law, where the GPT-2-like models continue to improve the next-token prediction huber loss with a scaling law of $\log_{10} L_{\text{pretraining}}(C) =-0.00448 \log_{10} C -2.386$. Such a scaling law also translates into a more robust and higher level of abstraction in the learned latent representation. The robust latent representation leads to a scaling law on the downstream inference MSE of the surface gravity of stars, with a scaling law of  $\log_{10}L_{\text{downstream}}(N) = -0.063\log_{10} N - 0.725$. With a labeled training set of 10,000 samples and a model size of $>85$ M parameters, the autoregressive models outperform the supervised Transformer model trained from scratch on 100,000 samples, and these results continue to extrapolate to a higher precision with up to $10^9$ parameters. At a fixed precision for $\log g$ inference, GPT-2 XL with $1.5$B parameters generative model can achieve such precision consistently with 3-10 times fewer labeled samples compared to the mere supervised learning approach.


In this study, we chose the GPT-2 model to demonstrate more principally the scalability of autoregressive generative models for astronomical time series data. Different parts of the models can no doubt be fine-tuned for better performance, both in terms of the choice of tokenization (in our case, simple MLP tokenization) and the context window (we limited to a very short context of 80 tokens). A longer context window, or following ideas from vision Transformers \cite{foundation_image}, the segmentation of light curves into patches and tokenizing individual patches might be a way forward. Further, here we are limited to the exploration of the scaling law at a fixed training set of 0.7B tokens; an additional ablation study on the training size will also be informative.

Nonetheless, the study demonstrates that the Transformer-based models that have shown wild success in real-world applications are also promising for astronomical time series data. The scalability of the models up to 1.5B parameters with our small training set vividly demonstrates the possibility of having a robust and generalizable foundational model for astronomical time series data with sufficient compute. Beyond asteroseismology, the same method can clearly also translate into other forms of astronomical time series data, such as understanding solar flares \cite{flare} or classifying and finding out-of-distribution transient events \cite{lsst}.

Ongoing and future surveys, TESS, ZTF, Rubin and SiTian, are estimated to collect observations at scales of 1T, 14T, 14P, and 100P, respectively. There will be a myriad of time series data, perhaps rivaling the text data on the internet and perhaps beyond. The scaling law motivates the dedication of compute resources to characterize astronomical time series data with these models. We forecast that the $\log g$ prediction MSE can be improved two folds than our GPT-2 XL model given $10^4$ PF-days, i.e., $\sim 10^5$ A100 GPU days. Looking forward, just as with much of the highly non-linear and complex data in real-life scenarios, while the still somewhat simple physical drivers of the light curves of celestial objects can be understood, to a certain extent, with analytical derivations and methods, robust inference and ultimate characterization might still require massive compute. The scalable model demonstrated here is a key step toward unlocking the full potential of astronomical time series data and paving the way for expediting discoveries in the ever-changing dark sky.

\section*{Acknowledgements}

We thank Mike Smith for constructive discussion and insights. YST acknowledges financial support received from the Australian Research Council via the DECRA Fellowship, grant number DE220101520. We also recognize the invaluable contribution of public data from the \textit{Kepler} mission, supplied by NASA's Ames Research Center. The \textit{Kepler} mission is funded by NASA's Science Mission Directorate. Our special thanks go to the \textit{Kepler} team for making their data openly accessible, thereby facilitating this research.

\bibliography{example_paper}
\bibliographystyle{icml2024}




\end{document}